\documentclass[12pt]{article}
\usepackage[english]{babel}
\usepackage[latin1]{inputenc}
\usepackage{times}
\usepackage[T1]{fontenc}
\usepackage{epsfig}
\usepackage{amsmath}
\usepackage{amssymb}

\begin{document}

\title{ Axial momentum  for the  relativistic  Majorana particle }

\author{H. Arod\'z  \\     
{\small  The Marian Smoluchowski Institute of Physics,
Jagiellonian University, Cracow, Poland \footnote{henryk.arodz@uj.edu.pl}} }

\date{$\;$}

\maketitle

\vspace*{1cm}

\begin{abstract}
The Hilbert space of states of the relativistic Majorana particle consists of normalizable bispinors with real components, and the usual momentum operator $- i \nabla$ can not be defined in this space. For this reason,  we introduce the axial momentum operator, $ - i \gamma_5 \nabla$ as a new observable for this particle.   In the Heisenberg picture, the axial momentum    contains a  component  which oscillates with the 
amplitude  proportional to $m/E$, where  $E$ is the energy and $m$  the mass of the particle.  The presence of the oscillations  discriminates between the massive and massless Majorana particle.   We show how the eigenvectors of the axial momentum, called the axial plane waves,  can be used as a basis for obtaining the general solution of the  evolution equation, also in the case of free Majorana field.  Here a novel feature is a coupling of  modes with the opposite momenta,  again present only in the case of massive particle or field.   
\end{abstract}

\pagebreak

\section{ Introduction}

 Recently  a lot of attention has been given  to  the Majorana particles and fields \cite{1},  both in  the context of  physics of neutrinos, see, e.g., \cite{2}, \cite{3}, and in condensed matter  physics,  \cite{4}, \cite{5}.  The theoretical framework used in these investigations  is  relativistic quantum field theory or  quantum statistical mechanics, respectively.  Historically, relativistic quantum field theory had been preceded by  relativistic quantum mechanics,  which still  continues to be one of the pillars  of the perturbative approach to scattering amplitudes in  quantum field theory.    Of course, the most popular examples of  relativistic quantum mechanics  suffer from such  well-known problems  as the presence of energies unbounded from below (in the case of  Dirac particle),  or  negative probability densities (for a scalar particle).  Nevertheless,  such quantum  mechanics  can be very useful.  Apart from the above mentioned background for the perturbative expansion,  on should mention also  its importance for  approximate description of many physical phenomena.   The importance of the Dirac equation for atomic physics is a good illustration of this point.

In our paper we address two interesting problems of  the relativistic quantum mechanics of a single free Majorana particle, which hitherto, to the best of our knowledge,  have not been discussed in literature.  The first problem is that in the Majorana case the standard momentum  $\hat{\mathbf{p}}= - i \nabla$ has to be replaced  with some other operator. The second problem arises from the fact  that the generator of time evolution, i.e., the Hamiltonian, is not Hermitian.   Therefore, according to the standard rules of quantum mechanics, it can not be accepted as the energy observable. Thus, we need a new energy operator.  Moreover,   it turns out that the  eigenvectors  of the Hamiltonian can not be  taken as a basis in the pertinent Hilbert space. Recall that  traditionally eigenvectors of the  Hamiltonian  form the basis in which we write the general solution of the wave equation,  irrespectively whether we consider the quantum mechanics or a free Majorana field.  Therefore, such  general solution   has to be written  in another basis.  
Below we offer solutions to all these problems.  On the whole, we present a self-contained formulation of the relativistic  quantum mechanics of  the Majorana particle  based on real bispinors.  

Let us stress that the same formalism 
is relevant also for  the free Majorana field. Nevertheless,  for the sake of  clarity,  below we use only the language of quantum mechanics.

 Let us  describe  the problems  mentioned above and  proposed solutions in more detail.   By definition, the Majorana bispinors \footnote{ For definitions, and a comparison with the Dirac and Weyl  bispinors, see, e.g.,  \cite{6}.} are invariant with respect to the charge conjugation operator $C$.
In the Majorana representation for the Dirac matrices $\gamma^{\mu}$, in which all these matrices are purely imaginary, the charge conjugation is reduced just to the complex conjugation.   Therefore, in this representation  all four components of the Majorana  bispinor  are real numbers.  (Let us note in passing that this fact is used in  \cite{7} to draw an analogy with the Boltzmann equation.) 
The pertinent Hilbert space  ${\cal H}$ consists of  all such real normalizable bispinors.  The ordinary momentum operator $\hat{\mathbf{p}}= - i \nabla$  turns  real bispinors  into complex ones,  therefore it is not an operator in ${\cal H}$.   The problem persists if we use a  non-Majorana representation for the Dirac matrices, because the momentum operator does not commute with the  charge conjugation $C$ in any representation.  
Below we introduce a  new operator, called by us the axial momentum and denoted as $\hat{\mathbf{p}}_5$, with the purpose to replace the ordinary momentum operator  $\hat{\mathbf{p}}$,  and we examine its properties. We find that there is an intriguing difference between massless and massive Majorana particles. In the latter case, the direction of  the axial momentum in the Heisenberg picture  oscillates in time,  see formulas (6) and (10)  below.   The amplitude of the oscillating component  is of the order  $m/ E$, where $m$ is the rest mass of the particle and $E$ its energy. In  the case of  neutrino with the  mass of the order $1 eV$ and the energy $1 MeV$  this amplitude is rather small, of the order $10^{-6}$.   As  the energy observable we adopt the operator $\hat{E} = \sqrt{m^2 + \hat{\mathbf{p}}_5^2}$  which does not have negative eigenvalues. 

The second  problem considered by us stems from the fact that   the Dirac equation  for the massive  Majorana  particle  is equivalent to an evolution equation with a Hamiltonian  which is  real and antisymmetric,   but it is not Hermitian.  Such a wave equation in quantum mechanics  is  not  standard one \footnote{But it is not  new,  see, e.g., \cite{8} and references therein, for other  non-Hermitian Hamiltonians.}. Therefore,  time evolution of  wave functions should be carefully examined, in particular, conservation of the norm.    We  check that the norm remains constant in time. Furthermore, the eigenvectors of the Hamiltonian have complex components, hence  they  can not be used as a basis in the  Hilbert space consisting of  real bispinors.   Instead, we  use the basis of  axial plane waves which are  the common eigenvectors of the axial momentum and of the new  energy operator.   Next, we find the general explicit solution of the wave equation.   Surprisingly, in the massive case the time evolution mixes the modes with the opposite values of the axial momentum.  The usual unitary factor $\exp(iEt)$ is replaced by two $SO(4)$ matrices of the form $\exp( K_{\pm} E t)$, see formulas (13).

It is worth mentioning  that in the presented below quantum mechanics of the Majorana particle we do not need any complex numbers.  They appear in our formulas, but  their presence is superficial -- the only reason for it is that we wish to adhere to the standard notation with the Dirac matrices. To some extent, the role of the imaginary unit $i$ is played by the real antisymmetric matrix $i\gamma_5$, which has the property $(i\gamma_5)^2 = -I$. We think that such explicit example of relativistic quantum mechanics devoid of  complex numbers can be interesting on its own right.

 The plan of our paper is as follows. In the next section  we
introduce the axial momentum,  as well as   other observables for the Majorana particle, including  the energy.   The time evolution of the axial momentum in the Heisenberg picture is investigated in Section 3.   Section 4  is devoted to the general solution of the wave equation. Section 5 contains a summary and several remarks.

\section{The axial momentum and other observables}
We adhere to the standard axiomatics of quantum mechanics.  In particular,  observables are represented by linear, Hermitian operators in a Hilbert space ${\cal H}. $     In our case,  elements of  ${\cal H} $ are real bispinors  $\psi(\mathbf{x}) = (\psi^{\alpha}(\mathbf{x}))$, $ \alpha =1,2,3,4,$  and the scalar product has the form  \[ \langle \psi_1|\psi_2 \rangle = \int \!  d^3x\: \psi_1^T(\mathbf{x}) \psi_2(\mathbf{x}), \] where $T$ denotes the matrix transposition. The bispinors are regarded as one-column matrices.  An observable $\hat{\cal{O}}$ obeys, in particular,  the condition $ \langle \psi_1|\hat{\cal{O}}\psi_2 \rangle =\langle \hat{\cal{O}}\psi_1|\psi_2 \rangle $ for all bispinors $\psi_1, \psi_2$ from the domain of $\hat{\cal{O}}$.  Furthermore,  the operator $\hat{\cal{O}}$ should be real, that is,  the bispinor $\hat{\cal{O}}\psi$  should be real like $\psi$. 
This last condition has the far reaching consequence: it eliminates the standard momentum operator  $\hat{\mathbf{p}} = - i \nabla$.  If we  just remove the imaginary unit $i$ to make it real,  the  $- \nabla$ operator is not  Hermitian.  

The time evolution of the real bispinors is governed by the Dirac equation, 
\[ i \gamma^{\mu} \partial_{\mu} \psi - m \psi =0,\]  where the Dirac matrices $\gamma^{\mu}$  are all purely imaginary, and 
$(\gamma^0)^T = -\gamma^0,\: (\gamma^k)^T = \gamma^k$, $(\gamma^0 \gamma^k)^T = \gamma^0 \gamma^k$, $\;k=1,2,3.$  A concrete choice for these matrices is made below formula (2).  
The Dirac equation can be rewritten as  
\begin{equation} 
\partial_t \psi = \hat{h} \psi, 
\end{equation}
where 
\[ \hat{h} = - \gamma^0 \gamma^k \partial_k - i m \gamma^0. \]
This operator is real, but it is not Hermitian.  Nevertheless, the scalar product  is constant in time, 
\[ \partial_t \int\!d^3x\: 
\psi_1^T(\mathbf{x}, t)\: \psi_2(\mathbf{x},  t)  = - \int\!d^3x\: \partial_k(\psi_1^T(\mathbf{x}, t)\:\gamma^0\gamma^k \psi_2(\mathbf{x},  t))  =   0\] 
if  $\psi_1(\mathbf{x}, t),  \psi_2(\mathbf{x}, t) $ obey Eq.\ (1) and  vanish sufficiently quickly at the spatial infinity. 

Hint about the form of the momentum observable comes from classical field theory. It is shown in \cite{9} that in the massless case ($m=0$),  Eq.\ (1) is equivalent to the Euler-Lagrange equation obtained from the Lagrangian  \[ {\cal L} = - i \overline{\psi} \gamma_5 \gamma^{\mu} \partial_{\mu} \psi,  \] 
where $\overline{\psi} = \psi^T \gamma^0, $ and $\gamma_5 = i \gamma^0  \gamma^1  \gamma^2  \gamma^3$.  The $ \gamma_5$  matrix is purely imaginary, $\gamma_5^2 = I, \; \gamma_5^T = - \gamma_5$.  Because $\psi$ is real, the presence of the $\gamma_5$ matrix is crucial, otherwise ${\cal L}$ would be a total divergence yielding no  evolution equation. The Noether theorem applied to this Lagrangian  yields the conserved 4-momentum $P^{\mu}$  of the classical Majorana field $\psi$ with real (not Grassmannian) components, namely
\[P^0= -i\int\!d^3x \: \psi^T \gamma^0 \gamma^k \gamma_5 \partial_k \psi, \;\;\;   P^k = -i\int\!d^3x \:\psi^T  \gamma_5 \partial_k \psi.    \]
These formulas  look like expectation  values of  the operators 
\[ \hat{E}_0 = \gamma^0 \gamma^k  \hat{p}^k_5,    \;\;\;  \hat{p}^k_5 = - i \gamma_5 \partial_k, \]
which are Hermitian, real, and they commute with each other.  The presence of $\gamma_5$ suggests the name for $\mathbf{p}_5= - i \gamma_5 \nabla$:  the axial momentum.  In spite of the fact that the Lagrangian ${\cal L}$ is for the massless Majorana field,  we consider these operators also when $m\neq 0$. 

The real matrix $i \gamma_5$ present in $ \hat{p}^k_5 $  has the property $(i \gamma_5)^2= -I$. Therefore it  may be regarded as a matrix replacement for the imaginary unit $i$ in $\hat{\mathbf{p}} = -i \nabla$. In this vein,  equation (1)  with $m=0$ can be written  in the equivalent form  
\[ i \gamma_5 \partial_t \psi = \hat{E}_0 \psi, \] which is the Schroedinger equation with $i$ replaced by the matrix $i \gamma_5$. Therefore, in the massless case $\hat{E}_0$ can be regarded as the Hamiltonian. Because it is Hermitian, one may adopt it as the energy operator, analogously as in the case of Dirac particle. As is well-known, such identification leads to  unphysical spectrum with    unbounded from below  negative energies. In the massive case, instead of $\hat{E}_0$ we have $i \gamma_5\hat{h}$. This last operator is not Hermitian (unless $m=0$), hence we have to seek  another energy operator.  The good candidate is  $\hat{E} = \sqrt{m^2 +\hat{\mathbf{p}}_5^2 }= \sqrt{m^2 -\Delta } $, which is the correct universal formula for the energy of arbitrary single, free relativistic particle, $\Delta$ denotes the Laplacian. This operator commutes with the Hamiltonian $\hat{h}$, hence its expectation values are constant in time. 
Moreover,   $\hat{E}^2 - \hat{\mathbf{p}}_5^2  = m^2,$ as expected for any single, free relativistic particle with the rest mass $m$ on the basis of Poincar\'e invariance in quantum field theory.   The spectrum of $\hat{E}$  is positive. There is no reason to introduce negative energies for the free Majorana particle  when we do not equate the energy operator with the generator of time evolution $\hat{h}$ (the Hamiltonian).  The energy operator $\hat{E}$ does not play any important role in our considerations because it is a simple function of $\hat{\mathbf{p}}_5$.

One can easily check that the operator $\hat{E}_0$ is proportional to the standard  helicity operator $\hat{\lambda} = \hat{\mathbf{S}}\:  \hat{\mathbf{p}}/ |\hat{\mathbf{p}}|$,  namely
\[ \hat{E}_0 = 2 |\hat{\mathbf{p}}| \hat{\lambda},  \] where  $\hat{\mathbf{p}} = -i \nabla$, 
$\: \hat{S}^j = i \epsilon_{jkl} [\gamma^k, \gamma^l] /8$ are the standard spin matrices, and $ |\hat{\mathbf{p}}| = \sqrt{ \hat{\mathbf{p}}^2}.  $    Notice that  $\hat{\mathbf{p}}^2 = - \nabla^2 =\hat{\mathbf{p}}_5^2, $  hence  $|\hat{\mathbf{p}_5}| = |\hat{\mathbf{p}}|$.  The operator  $\hat{\lambda}$ is real, as opposed to $\hat{\mathbf{S}}$ and $\hat{\mathbf{p}}$.  Thus, $\hat{E}_0$ is to be associated with the helicity rather than with the energy. Notice that the helicity operator can be written in the form $\hat{\lambda} = \hat{\mathbf{\Sigma}} \hat{\mathbf{p}}_5/|\hat{\mathbf{p}}_5|$, where $ \hat{\mathbf{\Sigma}} = \gamma_5 \hat{\mathbf{S}}$ is the correct spin matrix (real and Hermitian) in our formalism, see the Section 5. 

The axial momentum can be regarded as the generator of the spatial translations.  This can be inferred from the fact the  axial momentum is related to the Noether charge corresponding to the translations, as described above. One can also see this more directly. Let us start from the well-known formula for the translations which uses the standard momentum $\hat{\mathbf{p}} = -i\nabla$,  
\[ \exp(i \mathbf{a} \hat{\mathbf{p}}) \:\psi(\mathbf{x},t) = \exp( \mathbf{a} \nabla) \:\psi(\mathbf{x},t)=  \psi(\mathbf{x} + \mathbf{a}, t).    \] 
We see from this formula that  the translations are in fact generated by the operator $\nabla$. Because this operator is not Hermitian, one usually adds the coefficient $-i$ to $\nabla$ in order to produce a Hermitian operator (i.e., $\mathbf{p}$), and also the factor $i$ in the first exponent for correctness of the formula. In the Majorana case we simply use the real matrix $i \gamma_5$ instead of $i$.  Then, the formula for translations reads  
\[ \exp(i \gamma_5 \mathbf{a}\hat{\mathbf{p}}_5) \:\psi(\mathbf{x},t) = \exp( \mathbf{a} \nabla) \:\psi(\mathbf{x},t)=  \psi(\mathbf{x} + \mathbf{a}, t).    \] 
The exponential operator on the l.h.s. is, of course, unitary. The generators of rotations $\hat{\mathbf{\Sigma}}=\gamma_5 \hat{\mathbf{S}}$ are obtained in analogous manner.

Normalized eigenvectors of  the axial  momentum, which we call the axial plane waves,  are defined  by the conditions 
\[  \hat{\mathbf{p}}_5 \psi_{ \mathbf{p}}(\mathbf{x}) =  \mathbf{p}\:  \psi_{ \mathbf{p}}(\mathbf{x}),  \;\;\;\; \int\!d^3x\:  \psi^{T}_{ \mathbf{p}}(\mathbf{x})\: \psi_{\mathbf{q}}(\mathbf{x}) = \delta(\mathbf{p} - \mathbf{q}).  \] 
They have the form 
\begin{equation}
 \psi_{ \mathbf{p}}(\mathbf{x}) =  (2\pi)^{-3/2} \exp(i \gamma_5 \mathbf{p}\mathbf{x}) \: v,  
\end{equation}
where  $v$ an arbitrary real, constant, normalized ($ v^T v=1$) bispinor, and 
\[  \exp(i \gamma_5 \mathbf{p}\mathbf{x}) =  \cos(\mathbf{p}\mathbf{x}) I  + i \gamma_5 \sin( \mathbf{p}  \mathbf{x} ). \] 

In order to find the common eigenvectors of  $\hat{\mathbf{p}}$ and $\hat{E}_0$,  we choose for  the Dirac matrices
\[ \gamma^0 = \left(  \begin{array}{cc} 0 & \sigma_2 \\ \sigma_2 & 0 \end{array}  \right), \;\;    \gamma^1 = i \left(  \begin{array}{cc} - \sigma_0 & 0 \\ 0 & \sigma_0 \end{array}  \right), \;\;   \gamma^2 = i \left(  \begin{array}{cc} 0 & \sigma_1 \\ \sigma_1 & 0 \end{array}  \right), \;\;  \gamma^3 = -i \left(  \begin{array}{cc} 0 & \sigma_3 \\ \sigma_3 & 0 \end{array}  \right).   \]
Then,  
\[ \gamma_5 = i \left(  \begin{array}{cc} 0 & \sigma_0 \\ - \sigma_0 & 0 \end{array}  \right).  \]
Here $\sigma_k$ are the Pauli matrices, and $\sigma_0$ is the $2\times 2$ unit matrix. The condition 
\[  \hat{E}_0 \:  \psi_{ \mathbf{p}}(\mathbf{x})   =  E_0 \: \psi_{ \mathbf{p}}(\mathbf{x}) \]
leads to the matrix equation for the bispinor $v$
\[ \gamma^0 \gamma^k p^k \:v= E_0 \:v, \] 
well-known in the context of the Dirac equation.  The eigenvalues $E_0 = \pm |\mathbf{p}|$, which correspond to the helicities $\lambda = \pm 1/2$,  are double degenerate.   In the case
 $E_0= + |\mathbf{p}|$ we find
\begin{equation}
v_1^{(+)}(\mathbf{p}) = \frac{1}{\sqrt{2  |\mathbf{p}|  (  |\mathbf{p}|  - p^2) }} \left(\begin{array}{c} - p^3\\  p^2 - |\mathbf{p}| \\ p^1 \\ 0   \end{array}  \right),    \;\;\; v_2^{(+)}(\mathbf{p}) =  i \gamma_5\: v_1^{(+)}(\mathbf{p}), 
\end{equation}
and for $E_0 = - |\mathbf{p}|$ 
\begin{equation}  
v_1^{(-)}(\mathbf{p}) =  i \gamma^0\: v_1^{(+)}(\mathbf{p}),  \;\;\; v_2^{(-)}(\mathbf{p}) =  i \gamma_5\: v_1^{(-)}(\mathbf{p}) = - \gamma_5 \gamma^0 v_1^{(+)}(\mathbf{p}). 
\end{equation}
These bispinors are real, and orthonormal
\[  (v^{(\epsilon)}_j)^T(\mathbf{p}) \: v^{(\epsilon')}_k(\mathbf{p}) = \delta_{\epsilon \epsilon'} \delta_{jk}, \] 
where $\epsilon, \epsilon' = +, -$,  and $j, k = 1,2.$  Their concrete form depends on the choice for the Dirac matrices  $\gamma^{\mu}$, of course.  Notice that  the complete set (3), (4) of bispinors is generated from $v^{(+)}_1$ by acting with the antisymmetric real matrices $i \gamma_5,  \: i \gamma^0$.

\section{ Time evolution of the observables}
Formal solution of Eq.\ (1) has the form 
\[  | t \rangle = \exp(t \hat{h}) \:|t_0 \rangle,  \]   where  $|t_0\rangle$ denotes the initial state.  Because  $ \hat{h}^T = - \hat{h}$, the operator $ \exp(t \hat{h})$ is orthogonal one. Time dependent  expectation value of an observable $\hat{{\cal O}}$ has the form 
\[ \langle t |   \hat{{\cal O}} |t \rangle  = \:\langle t_ 0 | \exp(- t \hat{h}) \: \hat{{\cal O}}\:\exp(t \hat{h}) |t_0 \rangle .\]  Therefore,  
the Heisenberg picture version of  $\hat{{\cal O}}$  has the form 
\[  \hat{{\cal O}}(t) =  \exp(- t \hat{h})\: \hat{{\cal O}}\: \exp(t \hat{h}). \] 
It follows that 
\begin{equation}
\frac{d  \hat{{\cal O}}(t)}{ d t} = - \left[ \hat{h}, \:  \hat{{\cal O}}(t) \right] + (\partial_t  \hat{{\cal O}})(t),
\end{equation}
where the last term on the r.h.s. is present only if  $ \hat{{\cal O}}$ has an explicit time dependence in the Schroedinger picture -- in the present paper we do not consider such operators. 

The observables $\hat{E}_0, \; \hat{\mathbf{p}}^2, \; |\hat{\mathbf{p}}|$ and $\hat{\lambda}$ commute with $\hat{h}$, hence they are constant in time.  The axial momentum is constant in time only in the massless case.

 In the massive case,
\[  \left[ \hat{h}, \:\hat{p}_5^k \right] = -2 i m\: \gamma^0\: \hat{p}^k_5, \]  
hence
\[  
\frac{d  \hat{p}^k_5(t)}{ d t} = 2 i m \:\gamma^0(t) \: \hat{p}^k_5(t),  
\]

where
\[  \gamma^0(t) = \exp(- t \hat{h}) \gamma^0 \exp(t \hat{h}).\]
Because $[ \hat{h}, \nabla]=0$,  $\; \hat{p}^k_5(t)$  can be written in the form 
 \begin{equation}  \hat{p}^k_5(t) = - i \gamma_5(t) \partial_k, \end{equation}
  where $\gamma_5(t)$ obeys the equation 
\begin{equation}  \frac{d  \gamma_5(t)}{ d t} = 2 i m \:\gamma^0(t) \: \gamma_5(t). \end{equation}
The operator  $\gamma^0(t) \: \gamma_5(t)$ present on the r.h.s. of Eq.\ (7) obeys the following equation 
\begin{equation} \frac{d (\gamma^0(t) \: \gamma_5(t))}{dt} =  2 \gamma_5(t) \gamma^k(t) \partial_k + 2 i m \gamma_5(t), \end{equation}
and 
\[  \frac{d^2 (\gamma^0(t) \: \gamma_5(t))}{dt^2} = 4 (\partial_k \partial_k - m^2) \gamma^0(t) \gamma_5(t).
\]
 The latter equation has the following  solution 
\begin{equation} 
\gamma^0(t) \gamma_5(t) =  \cos(2 \hat{E}t) \gamma^0 \gamma_5 - i (\hat{E}_0 \gamma^0 - m \gamma_5) \hat{E}^{-1} \sin(2 \hat{E} t), 
\end{equation}
where $\hat{E} = \sqrt{m^2 - \partial_k \partial_k}$ is constant in time.  The solution (9) obeys the required initial conditions at $t=0$:  \[ 
\left. \gamma^0(t) \gamma_5(t) \right|_{t=0} = \gamma^0 \gamma_5, \;\;\;  \left. \frac{d (\gamma^0(t) \: \gamma_5(t))}{dt}\right|_{t=0} =  2 \gamma_5 \gamma^k \partial_k + 2 i m \gamma_5,    \] 
where the second condition follows from Eq.\  (8).   Next,  we insert  solution (9)  on the r.h.s. of Eq.\ (7) and integrate for  $\gamma_5(t)$, 
\begin{equation}  \gamma_5(t) = \gamma_5 + i m \hat{E}^{-1}\gamma^0 \gamma_5 \left[ \sin(2 \hat{E}t)  +    \hat{J}  (1- \cos(2 \hat{E}t))\right],  \end{equation}
where  $\hat{J} = \hat{h}/\hat{E}$.   Notice that   $\hat{J}^2 = - I.$  Therefore,  
the two oscillating terms present on the r.h.s of formula (10)  are of the same order  $m/\hat{E}$.  The absolute value of the axial momentum, $| \hat{\mathbf{p}} |= \sqrt{\hat{\mathbf{p}}^2} = \sqrt{- \nabla^2}$,  remains constant in time. 

One may  ask whether the axial momentum is the best replacement for the ordinary momentum, because it is not constant in time  in the case $m \neq0$.  We have searched for  other possibilities, without  satisfactory results.   In particular, we have considered the operators 
\[ \hat{p}_m ^k = \hat{p}_5^k - m \gamma_5 \gamma^k, \]
with $k=1, 2, 3.  $     They are real, Hermitian,  they commute with $\hat{h}$, and for  $m=0$ they coincide with  the axial momentum. Unfortunately, they do not commute with each other.

\section{Time evolution of the Majorana bispinor}
The axial plane waves can be used in a general  solution of the evolution equation (1).  Such solution is
  written as a  superposition of the axial plane waves (2),  which appear in place of the   ordinary plane waves.   Thus, the  
expansion  of $\psi$  into the axial plane waves has the form 
\begin{equation}\psi(\mathbf{x}, t) = (2\pi)^{-3/2} \sum_{\alpha=1}^2 \int\!d^3p\: e^{i \gamma_5 \mathbf{p} \mathbf{x}} \left(v_{\alpha}^{( +)}(\mathbf{p})  c_{\alpha}(\mathbf{p},t) +  v_{\alpha}^{( -)}(\mathbf{p})  d_{\alpha}(\mathbf{p},t)\right). \end{equation}
 Equation (1)  gives the following  equations for the real amplitudes $c_{\alpha}, \:d_{\alpha}$:
\[ E_p^{-1} \: \dot{c}_1(\mathbf{p}, t) = n^2 c_2(\mathbf{p}, t)  + n^1 c_1(-\mathbf{p}, t) -  n^3 c_2(-\mathbf{p}, t), \] \[ E_p^{-1}\: \dot{c}_2(\mathbf{p}, t) = - n^2 c_1(\mathbf{p}, t)  - n^3 c_1(-\mathbf{p}, t) -  n^1 c_2(-\mathbf{p}, t),    \]
\[ E_p^{-1}\: \dot{d}_1(\mathbf{p}, t) = - n^2 d_2(\mathbf{p}, t)  + n^1 d_1(-\mathbf{p}, t) + n^3 d_2(-\mathbf{p}, t), \]
\[ E_p^{-1}\: \dot{d}_2(\mathbf{p}, t) =  n^2 d_1(\mathbf{p}, t)  + n^3 d_1(-\mathbf{p}, t) -  n^1 d_2(-\mathbf{p}, t), \]
where the dots stand for $d/dt, \:$  $ E_p = \sqrt{\mathbf{p}^2 + m^2} $, and 
\[  n^1= \frac{m\: p^1}{E_p \:\sqrt{(p^1)^2 + (p^3)^2}}, \;\; n^2 =  \frac{|\mathbf{p}|}{E_p}, \;\;  n^3= \frac{m \: p^3}{E_p\: \sqrt{(p^1)^2 + (p^3)^2}}.    \]
The coefficients $n^1, n^2, n^3$  are proportional to the scalar products  $v_{\alpha \epsilon}^T\: v_{\beta \epsilon'}$.   Note that  $\mathbf{n}^2 =1$.  

We see that   in the case $m \neq 0$ rather unexpected mixing  between the modes with the opposite eigenvalues  $\mathbf{p}$ and $-\mathbf{p}$ of the axial momentum appears. 
 Looking back at Eq.\ (1),  the coupling of the modes $\mathbf{p}$ and $ - \mathbf{p}$ appears because  $\gamma^0 \exp(i \gamma_5 \mathbf{p} \mathbf{x})$ $ = $ $ \exp( - i \gamma_5 \mathbf{p} \mathbf{x}) \gamma^0$.  It  is present always when $(n^1)^2 + ( n^3)^2 > 0$,  but it can be  rather  weak.  For example,  if we take   $m= 1$~~eV and $E_p = 1$ MeV,  then $\sqrt{(n^1)^2 + ( n^3)^2}= m/E_p = 10^{-6}  $ and $n^2= \sqrt{1- m^2/E_p^2}\approx 1$.  On the other hand,  the mixing is dominant when  $E_p \approx m$.

In order to solve  the equations for the amplitudes $c_{\alpha}, d_{\alpha}$,  we split the amplitudes into the even and odd parts, 
\[ c_{\alpha}(\mathbf{p},t) =  c_{\alpha}^{\:'}(\mathbf{p},t) +  c_{\alpha}^{\:''}(\mathbf{p},t),  \;\;\; d_{\alpha}(\mathbf{p},t) =  d_{\alpha}^{\:'}(\mathbf{p},t) +  d_{\alpha}^{\:''}(\mathbf{p},t),   \]
where $c_{\alpha}^{\:'}(- \mathbf{p},t) = c_{\alpha}^{\:'}(\mathbf{p},t), $ $\;c_{\alpha}^{\:''}(- \mathbf{p},t) =- c_{\alpha}^{\:''}(\mathbf{p},t)$, and similarly for $d', d''$. Such a splitting is unique.  Using the notation 
\[ \vec{c}(\mathbf{p},t) =  \left(\begin{array}{c} c_1' \\ c_1''\\ c_2'\\ c_2'' \end{array} \right)\!\!, \:
 \vec{d}(\mathbf{p},t) =  \left(\begin{array}{c} d_1' \\ d_1''\\ d_2'\\ d_2'' \end{array} \right)\!\!,\:  K_{\pm}(\mathbf{p}) = \left(\begin{array}{cccc} 0 & -n^1& \pm n^2&\pm n^3\\ n^1& 0 &\mp n^3 &\pm n^2\\ \mp n^2& \pm n^3 &0& n^1 \\ \mp n^3 & \mp n^2 & - n^1 &0 \end{array}\right)\!\!,  \] we rewrite the equations for the amplitudes in the form
\begin{equation} \dot{\vec{c}}(\mathbf{p},t) =   E_p \: K_+(\mathbf{p})  \:   \vec{c}(\mathbf{p},t),   \;\;\;     \dot{\vec{d}}(\mathbf{p},t) =  E_p\: K_-(\mathbf{p}) \:   \vec{d}(\mathbf{p},t).    \end{equation}
Solutions of these equations have the form 
\begin{equation} \vec{c}(\mathbf{p},t) =  \exp(E_p \: K_+(\mathbf{p})\:t)  \:    \vec{c}(\mathbf{p},0),   \;\; \vec{d}(\mathbf{p},t) = \exp( E_p\: K_-(\mathbf{p})\: t)  \:   \vec{d}(\mathbf{p},0).    \end{equation}
The  matrices $K_{\pm}$ are antisymmetric,  hence the exponential matrices present in (13) are orthogonal ones -- they belong to the $SO(4)$ group.   Furthermore,  because $K_{\pm}^2 = - I$,  we may write 
\begin{equation} \exp( E_p\: K_{\pm}(\mathbf{p})\: t) =  \cos(E_p\:t) I + \sin( E_p\:t) K_{\pm}(\mathbf{p}).  \end{equation}

The fact that the general solution (13) has such a simple form is a nice surprise.  Because the axial plane waves  are eigenvectors of  the operator $\hat{\mathbf{p}}_5$  which does not commute with $\hat{h}$ 
when $m \neq 0$, one could expect much more complicated time evolution of the amplitudes.

\section{Summary and remarks}

\noindent {\bf 1.}  We have presented a self-contained  formulation of the relativistic quantum mechanics of the single  free Majorana particle.  As one of the main results, we have shown that the axial momentum operator  $\mathbf{p}_5= -i \gamma_5 \mathbf{\nabla}$  is a viable  replacement for the ordinary momentum  $\mathbf{p} = -i \mathbf{\nabla}$. In particular, it can be regarded as the generator of spatial translations.   Its  eigenfunctions  -- the axial plane waves -- provide a new basis  for mode decomposition of  Majorana bispinors.  In this basis, the time evolution is represented by  $SO(4)$ matrices (14), which replace the standard complex exponents $\exp(\pm i E_p t)$ -- this is our second main result. The energy operator $\hat{E} = \sqrt{m^2 + \hat{\mathbf{p}}_5^2}$ differs from the Hamiltonian $\hat{h}$ (which is not Hermitian), but it coincides with the well-known formula for the energy of a free relativistic particle.   

There are intriguing effects  due to  nonvanishing rest mass of the particle.  First, the direction of the axial momentum oscillates  if $m\neq0$ (the modulus remains constant). One may regard these oscillations as a kind of Zitterbewegung in the momentum space. In the well-known case of the Dirac particle, the Zitterbewegung is present in the position space -- the velocity operator $d \hat{\mathbf{x}}/dt$ is not constant. 
Second, there is the coupling  between  the modes  with the opposite values of the axial momentum. The strength of these effects is proportional to $m/E_p$, where $E_p= \sqrt{m^2 + \mathbf{p}_5^2}$ is the energy of the particle. \\

\noindent {\bf 2.}  The Dirac equation for the real Majorana bispinor $\psi(\mathbf{x}, t)$ of course possesses the Poincar\'e symmetry, as well as  the $P$ and $T$ symmetries.  The Poincar\'e  transformations of the wave function have the standard form,  \[ \psi'_{L,a}(x) = S(L) \psi(L^{-1}(x-a)), \]
where $x= (t,  \mathbf{x}), \;$ $S(L) = \exp(\omega_{\mu\nu} [\gamma^{\mu}, \gamma^{\nu}]/8)$, and $\omega_{\mu\nu}$ parameterize the proper orthochronous Lorentz group. Starting from this formula, one may obtain in the standard manner the generators of rotations, which are related to the spin  observable. The only difference is that the coefficients $i$ used in order to obtain Hermitian matrices in the Dirac case, here should be replaced by $i\gamma_5$, analogously as in the case of momentum.  

The $P$ and $T$ transformations have the form, respectively, 
\[ \psi'_P(\mathbf{x},t) = \eta_P i \gamma^0 \psi(-\mathbf{x}, t), \;\;\; \psi'_T(\mathbf{x},t) = \eta_T  \gamma_5 \gamma^0 \psi(\mathbf{x}, - t),\]
where  $\eta_p= \pm1, \eta_T = \pm 1$ are  intrinsic  parities of the particle.  By  definition,   the symmetry  transformations  applied to an arbitrary solution of  the Dirac equation yield   solutions of the same equation.  \\

\noindent {\bf 3.} The results of present work suggest several interesting topics for further research. First, one may delve deeper into the quantum mechanics  of the Majorana particle. This includes investigations of evolution of wave packets in external potentials  representing interactions with other particles. Furthermore,  one could   propose concrete procedures for measuring the axial momentum.      One can also  look for applications in particle physics,  and in condensed matter physics.  

Second, the axial momentum is the observable that can be used also in the case of Dirac particle, which is described by a complex bispinor. Such a reformulation of  the relativistic quantum mechanics of the Dirac particle with the axial momentum in place of the standard momentum may provide an interesting complementary  viewpoint on the Dirac particle.   

Last but not least, the mode decomposition (11)  with the axial plane waves  can be  used as the starting point  for quantization of the Majorana and the Dirac fields.  Here the interesting question is whether the resulting Fock spaces of free particles are equivalent to the standard ones.    Work in this direction is in progress. 

 \section{Acknowledgement}

We would like to thank Professor Kacper Zalewski for an inspiring discussion.

\end{document}